\documentclass[trackchanges,twocolumn,tight]{aastex701}
\usepackage{amsmath,graphicx,array}
\usepackage{dcolumn,soul}%
\usepackage{amsthm}
\usepackage{bm}
\usepackage[figuresright]{rotating}%
\usepackage{algorithm, algorithmicx, algpseudocode}
\usepackage{listings}%
\usepackage{hyperref}

\newcommand{\Referee}[1]{#1}
\begin{document}

\title{Rigorous analytic solution to the gravitational-wave overlapping event
rates}

\author[0000-0002-8742-8397]{Ziming Wang}
\affiliation{Department of Astronomy, School of Physics, Peking University, 
Beijing 100871, China}
\affiliation{Kavli Institute for Astronomy and Astrophysics, Peking University, 
Beijing 100871, China}
\email{zwang@pku.edu.cn}

\author[0000-0002-3081-0659]{Zexin Hu}
\affiliation{Department of Astronomy, School of Physics, Peking University, 
Beijing 100871, China}
\affiliation{Kavli Institute for Astronomy and Astrophysics, Peking University, 
Beijing 100871, China}
\affiliation{Theoretical Astrophysics, Eberhard Karls University of T\"{u}bingen,
T\"{u}bingen 72076, Germany}
\email{huzexin@pku.edu.cn}

\author[0000-0002-1334-8853]{Lijing Shao}
\affiliation{Kavli Institute for Astronomy and Astrophysics, Peking University, 
Beijing 100871, China}
\affiliation{National Astronomical Observatories, Chinese Academy of Sciences, 
Beijing 100012, China}
\email[show]{lshao@pku.edu.cn}
\correspondingauthor{Lijing Shao}

\begin{abstract}
In the era of the next-generation gravitational-wave detectors, signal overlaps
will become prevalent due to high detection rate and long signal duration,
posing significant challenges to data analysis. While effective algorithms are
being developed, there still lacks an integrated understanding on the
statistical properties for the population of overlapping
compact-binary-coalescence signals. For the first time, in order to aid rapid
and robust estimation, we rigorously derive and establish analytical expressions
for the expectation and variance for the number of overlapping events.
This framework is highly extensible, allowing analytical calculation
for more complicated scenarios, such as multi-signal overlaps, overlaps between
different types of sources, and source-dependent thresholds. We also
mathematically prove that the time difference between events in a single
observation run is described by the beta distribution, offering an analytical
prior reference for Bayesian analysis.  
\end{abstract}

\keywords{\uat{Gravitational waves}{678} --- \uat{Astrostatistics}{1882} --- \uat{Astrostatistics distributions}{1884}}

\section{Introduction}

\label{sec:intro}

Since the first direct detection of gravitational waves (GWs) by the Advanced
LIGO in 2015~\citep{Abbott:2016blz}, the LIGO-Virgo-KAGRA Collaboration has
detected about one hundred events from compact binary coalescence
(CBC)~\citep{LIGOScientific:2018mvr, LIGOScientific:2020ibl,
LIGOScientific:2021djp, LIGOScientific:2021usb}, which opens a new window to
explore important questions in fundamental physics, astrophysics, and
cosmology~\citep{LIGOScientific:2016lio, LIGOScientific:2017adf,
LIGOScientific:2018cki, LIGOScientific:2018dkp, LIGOScientific:2019fpa,
LIGOScientific:2020kqk, LIGOScientific:2021sio, KAGRA:2021duu}. Currently, the
rate of GW detections is several events per
week~\citep{LIGOScientific:2020ibl,LIGOScientific:2021djp, LIGO2024}.
\Referee{The next-generation (XG) ground-based GW detectors, such as the Cosmic
Explorer~\citep[CE; ][]{Reitze:2019iox, Reitze:2019dyk} and the Einstein
Telescope~\citep[ET; ][]{Punturo:2010zz, Hild:2010id, Sathyaprakash:2012jk,
Abac:2025saz}, are under development~\citep{LIGOScientific:2016wof,
Maggiore:2019uih}}.  They have an order of magnitude higher sensitivity and a
wider accessible frequency band compared to current ones. In the new era of
CE/ET, there will be many more and longer GW signals, $\sim 10^5$ CBC events per
year with effective duration from hours to days, and therefore signal overlaps
naturally arise~\citep{Regimbau:2009rk, Samajdar:2021egv, Himemoto:2021ukb,
Relton:2021cax, Pizzati:2021apa, Johnson:2024foj}. \Referee{The overlapping
signals also exist in the near-future space-borne GW detectors, such as the
Laser Interferometer Space Antenna~\citep[LISA; ][]{LISA:2017pwj},
Taiji~\citep{Hu:2017mde} and TianQin~\citep{TianQin:2015yph, Gong:2021gvw}
programs, where the stellar-mass CBC signals can last for months to years.}

Inappropriate modeling or analysis of overlapping signals will bias the
inference of source parameters and further bias astrophysical
implications~\citep{Pizzati:2021apa, Samajdar:2021egv, Relton:2021cax,
Hu:2022bji, Wu:2022pyg, Wang:2023ldq, Dang:2023xkj}. Besides the efforts in
developing efficient algorithms for identification as well as unbiased method
for parameter estimation of overlapping signals~\citep{Samajdar:2021egv,
Himemoto:2021ukb, Relton:2021cax, Langendorff:2022fzq, Janquart:2022fzz,
Alvey:2023naa, Miller:2023rnn, Papalini:2025exy, Hu:2025vlp, Baka:2025yqx},
another relevant and important question is on the population properties of
overlapping events, such as the detection rate. Currently in most
studies~\citep{Samajdar:2021egv, Himemoto:2021ukb, Hu:2022bji}, the number of
overlapping event is estimated by simulations, while there exist some analytical
expressions only for estimating the expectation of the overlapping-event
number~\citep{Pizzati:2021apa, Johnson:2024foj}. The variety of definitions for
overlapping events, such as time chunks with more than one
signal~\citep{Samajdar:2021egv, Pizzati:2021apa}, time-frequency
crossings~\citep{Johnson:2024foj}, or effectively the parameter-estimation
biases due to overlapping~\citep{Wang:2023ldq, Dang:2023xkj}, also limits a more
in-depth discussion.

In this work, we adopt a simple and easily extendable definition: an overlapping
event occurs whenever the time difference between adjacent events is less than a
threshold $\Delta t_{\rm th}$.  For the first time, we develop a systematic
study of overlapping-event population by rigorously deriving the statistical
properties of the overlapping-event number, which consist of the distribution
function,and the expectation as well as variance for a given total event number
$n$ and observation duration $\tau$. Based on our analytical results, we discuss
the validity of the binomial and Poisson approximations in the era of XG GW
detectors.  The expectation and variance after marginalizing $n$, and their
asymptotic expressions for a large detection rate $r$ and small normalized
threshold $\Delta t_{\rm th}/\tau$, are given in concise forms, depending only
on two expected event numbers, $\lambda := r\tau$ and $\epsilon := r\Delta
t_{\rm th}$.  \Referee{Then, we consider the cases of multiple-signal overlaps, 
overlaps between different types of sources, and source-dependent thresholds,
and all of them can be rigorously derived in our framework with straightforward
extensions.} We also prove that the distribution of the time differences between
events in a single observation run is on average the beta distribution, which
highly agrees with the simulation-based results~\citep{Himemoto:2021ukb}. These
analytical results provide a robust theoretical framework and precise analytical
tools for understanding overlapping-event populations, which is expected to
benefit the development of search and parameter-estimation algorithms for
overlapping signals in the community.

This paper is organized as follows. In Section~\ref{sec:assumptions}, we
introduce the model assumptions and conventions in this work. In
Section~\ref{sec:distribution of S}, we analytically derive the distribution of
the overlapping-event number given the total event number $n$, and calculate its
expectation and variance in both fixed and marginalized $n$ cases. The
asymptotic expressions of the expectation and variance, and the approximations
with the binomial and Poisson distributions are also discussed. \Referee{In
Section~\ref{sec:extensions}, we extend the results for two-signal overlaps with
a fixed threshold to more complex cases, including three or more signal
overlaps, overlaps between different types of sources, and source-dependent
thresholds.} In Section~\ref{sec:distribution of time differences}, we discuss
the distribution of time differences in a single observation run. Finally, we
summarize our results in Section~\ref{sec:conclusions}. For reader's
convenience, most of the derivation details are collected in
Appendix~\ref{sec:appendix}, and in the main text we focus on the representation
and discussion of the results.

\section{Assumptions and Conventions}\label{sec:assumptions}

Within a given time period $[0,\tau]$, the number of detected events, $N$, is a
random variable and follows a Poisson distribution ${\rm Pois}(r\tau)$ under a
constant detection rate $r$. We assume that every event independently occurs
with equal probability in the interval $[0,\tau]$. The arrival times of these
events are recorded sequentially in a GW detector\footnote{For a detector
network, the arrival times are defined in the earth-centered frame.},
corresponding to $n$ random variables $\big\{T_i\big\}_{i=1}^n$, where $0\leq
T_1\leq T_2\leq \cdots \leq T_n \leq \tau$. The above assumptions are accepted
in most studies estimating the overlapping event rate~\citep{Samajdar:2021egv,
Himemoto:2021ukb, Pizzati:2021apa, Johnson:2024foj}.

In this work, we define that an overlapping event occurs whenever the time
difference between adjacent events ${\Delta T}_i = T_i - T_{i-1} \,(i =
2,\cdots,n)$, is less than a fixed threshold ${\Delta t}_{\rm th}$.  This
definition for overlapping events is simple and easy to extend---for example, to
consider ${\Delta t}_{\rm th}$'s dependence on source
parameters~\citep{Samajdar:2021egv, Himemoto:2021ukb, Relton:2021cax}. We only
count the two-signal overlapping, that is, ${\Delta T}_i \leq {\Delta t}_{\rm
th}$ and ${\Delta T}_{i+1} \leq {\Delta t}_{\rm th}$ are considered as two
overlapping events, regardless of whether ${\Delta T}_i + {\Delta T}_{i+1} \leq
{\Delta t}_{\rm th}$. \Referee{The number of overlapping events can be expressed
as $S = \sum_{i=2}^n I_i$ with the overlapping variable $I_i =1$ if ${\Delta
T}_i \leq {\Delta t}_{\rm th}$ and $0$ otherwise.} In Section~\ref{sec:three or
more events overlap} we calculate the numbers of three or more signal overlaps,
and find that they are much smaller than the two-signal overlapping number as
expected.

For mathematical convenience, we define the dimensionless variables ${X}_i =
{T}_i/\tau$, ${\Delta X}_i = {\Delta T}_i/\tau$ and $\xi = {\Delta t}_{\rm
th}/\tau$. We also denote $\Delta T_1 = T_1$, and use bold letters to represent
the collection of these variables, such as ${\bm X} \equiv \{X_i\}_{i=1}^n$.
Throughout this work, random variables are denoted by uppercase letters, such as
$N$, $S$, $T_i$ and $X_i$, while their specific realizations are denoted by
lowercase letters, $n$, $s$, $t_i$ and $x_i$. Besides, we use $P$ to represent
distribution function and ${\rm Pr}$ for probability. 

Now we rephrase our model in a more mathematical form. After normalizing with
the observation duration $\tau$, the dimensionless times, ${\bm  X}$, are the
order statistics of $n$ independent and identically distributed (i.i.d.)\ random
variables from the uniform distribution ${\cal U}(0,1)$, and $S$ is the number
of adjacent order statistics that have time difference less than $\xi$.
\Referee{Based on the properties of the order
statistics~\citep{casella2024statistical}, the joint distribution of $\Delta
{\bm X}$ is}
\begin{equation}
    P(\Delta {\bm  x}) = n!\,,\ \  {{\Delta x}}_i\geq 0\ \  \& \ \  {\sum
    \nolimits}_{i=1}^n {{\Delta x}}_i\leq 1\,,
    \label{eq:joint distribution of dimensionless time difference}
\end{equation}
which is the basis for following calculations. \Referee{For readers'
convenience, we review the derivation of this distribution and its properties in
Appendix~\ref{appsub:joint and marginal distributions of time differences}.}

\section{Distribution, Expectation and Variance of the Overlapping-Event Number}
\label{sec:distribution of S}

\subsection{Exact Expressions}\label{sec:exact expressions}

We first calculate the number of overlapping events with a known event number,
$N = n$. Defining event $A_i = \big\{I_i =1\big\}$ and its complement
$\bar{A}_i$, we express the probability of detecting $s$ overlapping events in
$A_i$ and $\bar{A}_i$,
\begin{align}
    P(s|n) = C_{n-1}^s {\rm Pr}\left(\big (\cap_{i=2}^{s+1} {A}_i\big)\cap
    \big (\cap_{i=s+2}^{n} \bar{A_i}\big)\right)\,,\label{eq:intersection of A_i}
\end{align}
\Referee{where $C_{a}^b = {a!}/[{b!(a-b)!}]$ denotes the binomial coefficient.}
According to Eq.~\eqref{eq:joint distribution of dimensionless time difference},
we find that 
\begin{equation}
    {\rm Pr}\, \Big(\big\{\Delta X_1>\xi_1,\cdots,\Delta X_m>\xi_m\big\}\Big) =
    \left(1-\sum_{i=1}^{m}\xi_i\right)^n\,,
    \label{eq:prob of dimensionless time difference}
\end{equation}
for $\sum_{i=1}^{m}\xi_i \leq 1$. Then, the intersection of any $m$ events
among $\big\{\bar{A}_i\big\}_{i=2}^n$ is given by $(1-m\xi)^n$. With the help of
inclusion-exclusion principle, we further express the intersection of $A_i$ in
the form of intersection of $\bar{A}_i$ (see in Appendix~\ref{appsub:derive details of S distribution}) and calculate $P(s|n)$ as
\begin{equation}
    P(s|n) = C_{n-1}^s\sum_{k=0}^s C_{s}^k(-1)^{k}\Big[1-(n+k-s-1)\xi \Big]^n\,.
    \label{eq:prob of S explicit}
\end{equation}

It is more useful and intuitive to give the expectation and variance of $S$ when
estimating the overlapping-event rates. We find that they can be calculated
without the complex expression of $P(s|n)$. The expectation of $S$ given $n$
reads
\begin{equation}
    {\rm E}[S|n] = \sum_{i=2}^n {\rm E}[I_i|n] = (n-1)\Big[1-(1-\xi)^ n\Big]\,,
    \label{eq:expectation of S}
\end{equation}
which is based on the additive property of the expectation and ${\rm E}[I_i|n] =
1-{\rm Pr}(\bar{A}_i) = 1-(1-\xi)^n$.  For the variance, we need to calculate
the covariance between $I_i$ and $I_j$ for $i\neq j$. This is found to be ${\rm
Cov}(I_i,I_j) = (1-2\xi)^n - (1-\xi)^{2n}$ in Appendix~\ref{appsub:expectation and variance of S}.  It is worth
noting that $I_i$ and $I_j$ are negatively correlated, ${\rm Cov}(I_i,I_j)<0$ when $0 <\xi <1/2$. This is consistent with our intuition that if one time
difference between two events is large, other time differences are more likely
to be small under the constraint $\sum_{i=1}^n {\Delta T}_i \leq \tau$. The
variance of $S$ given $n$ reads 
\begin{align}
    {\rm Var}[S|n] = (n-1)\Big[&(1-\xi)^n+(n-2)(1-2\xi)^n \notag \\
    &- (n-1)(1-\xi)^{2n}\Big]\,.    \label{eq:variance of S}
\end{align}

Given the conditional expectation and variance of $S$ in
Eqs.~(\ref{eq:expectation of S}--\ref{eq:variance of S}), we now combine them
with the distribution of $N$. It involves some lengthy summations with Poisson
distribution, which are simplified in Appendix~\ref{appsub:expectation and variance of S}. The final
expressions are
\begin{align}
    {\rm E}[S] &=(\lambda - 1)(1 - e^{-\epsilon}) + \epsilon
    e^{-\epsilon}\,,\label{eq:expectation and variance combined with N} \\
    {\rm Var}[S] &= {\rm E}[S] +e^{-2 \epsilon} \Big[1 + e^{2 \epsilon} +
    \epsilon (2 + 3 \epsilon - 2 \lambda)\notag
    \\ 
    &\phantom{{}={\rm E}[S] +e^{-2 \epsilon} \Big[} - 2 e^{\epsilon} (1 +
    \epsilon + \epsilon^2 - \epsilon \lambda)\Big]\,, \notag
\end{align}
where we have introduced two dimensionless quantities,
 $\lambda  := r\tau$ and $\epsilon := \lambda \xi =
r{\Delta t}_{\rm th}$, whose physical meanings are clearly the expected
number of events in the observing duration and the overlapping threshold,
respectively.

\subsection{Asymptotic Behaviors}\label{sec:asymptotic expressions}

The above formulae are exact but somehow lengthy. In practice, they can be
further simplified to a more intuitive form after taking into account the
realistic values of the two dimensionless parameters, $\lambda$ and $\epsilon$.
Considering $\lambda$, the currently inferred merger rates for binary black
holes (BBHs) and binary neutron stars (BNSs) are ${\cal O}(10) \, {\rm
Gpc}^{-3}\,{\rm yr}^{-1}$ and ${\cal O}(100) \, {\rm Gpc}^{-3}\,{\rm yr}^{-1}$,
respectively~\citep{LIGOScientific:2020kqk, KAGRA:2021duu}. \Referee{These correspond to
detection rates of ${\cal O}(10^4-10^5) \,{\rm yr}^{-1}$ for BBHs and ${\cal
O}(10^5-10^6)\, {\rm yr}^{-1}$ for BNSs in the XG-detector
network~\citep{Samajdar:2021egv, Himemoto:2021ukb, Pizzati:2021apa, Hu:2022bji}.} The
uncertainties of the detection rates are within an order of magnitude across
different population models.  Considering $\epsilon$, the criterion for
overlapping events is in debate.  As mentioned, the overlapping events could be 
defined as those signals that significantly affect the parameter estimation of
each other~\citep{Wang:2023ldq, Dang:2023xkj, Johnson:2024foj}, and one should
consider its source-dependence~\citep{Samajdar:2021egv, Himemoto:2021ukb,
Relton:2021cax}. Frequency-evolution crossing serves as another possible
criterion~\citep{Johnson:2024foj}. \Referee{Roughly speaking, in the XG detectors'
    data streams, the thresholds for BBH and
BNS overlapping events were estimated to be ${\Delta t}_{\rm th} \sim 0.1\,{\rm s}$ and ${\Delta
t}_{\rm th} \sim 1\,{\rm s}$,
respectively~\citep{Samajdar:2021egv, Relton:2021cax, Pizzati:2021apa,Antonelli:2021vwg}.
Based on these studies, for a one-year observing duration we choose $\lambda = 10^5$
and $\epsilon = 10^{-3}$ as representative values in the XG detection era.
However, if the detection rate is estimated optimistically and the threshold is
chosen to be more relaxed---for example, one 
order of magnitude larger---these values will be $\lambda = 10^6$ and $\epsilon =
10^{-1}$, which can be regarded as an extreme case for the upper bound of the
overlapping-event number.} Besides, when $\lambda$ is large, the observed 
total event number approaches $\lambda$ with a variance of $\sqrt{\lambda}$, 
thus we have $\epsilon_n := n\xi \sim \epsilon$. Recalling that $\xi = {\Delta
t}_{\rm th}/\tau$, $\epsilon_n$ can be interpreted as the ratio of the time
occupied by $n$ overlapping-windows to the total observation duration $\tau$.
Given the above discussion, in the following we study the behavior of
Eqs.~(\ref{eq:expectation of S}--\ref{eq:expectation and variance combined
with N}) in the limit of $\lambda,n\gg 1$ and $\epsilon,\epsilon_n \ll 1$.

The asymptotic expressions of the expectation
and variance of $S$, given $n$, are
\begin{equation}
	\begin{aligned}
        {\rm E}[S|n] &= (n-1)\epsilon_n+{\cal O}(n\epsilon_n^2)\,, \\
		{\rm Var}[S|n] &= (n-1)\epsilon_n+{\cal O}(n\epsilon_n^2)\,,
    \end{aligned}
      \label{eq:asymptotic fix n}
\end{equation}
where the leading terms are both $(n-1)\epsilon_n$. 
To test these approximations, we calculate the expectation and variance of $S$
according to the exact expression Eq.~\eqref{eq:expectation and variance combined
with N} and the approximation Eq.~\eqref{eq:asymptotic fix n}. Taking $n=10^5$ and
$\epsilon_n = 10^{-3}$, we find that ${\rm E}[S|n] = 99.94$ ($100.00$) and
$\sigma\equiv \sqrt{V[S|n]} = 9.987$~($10.000$) according to the exact
(approximate) expression. The relative errors of $0.06\%$ and $0.13\%$ are
consistent with the ${\cal O}(\epsilon_n)$ relative correction from the
next-to-leading order. For the extreme case of $n= 10^6$ and $\epsilon_n
= 0.1$, we have ${\rm E}[S|n] = 9.52\times 10^4$ ($1.00\times 10^5$) and $\sigma = 2.79 \times 10^2$ ($3.16 \times 10^2$),
corresponding to $5.1\%$ and $13.3\%$ relative errors, respectively.

The asymptotic expressions of the expectation and variance of $S$ after
marginalizing $n$ are
\begin{equation}
	\begin{aligned}
        {\rm E}[S] &= \lambda\epsilon+{\cal O}(\lambda\epsilon^2)\,,\\
        {\rm Var}[S] &=  \lambda\epsilon+{\cal
        O}(\lambda\epsilon^2)\,,
    \end{aligned}
    \label{eq:asymptotic marginalizing n}
\end{equation}
where one once again notices the leading terms equal to each other.
The
similarity between Eqs.~\eqref{eq:asymptotic fix n} and \eqref{eq:asymptotic
marginalizing n} can be explained by the fact that, for a large $\lambda$, the
Poisson random variable $N$ is almost always around $\lambda$. \Referee{It is worth
mentioning that the corrections from the next-to-leading order of the expectation
are found to be
negative for both fixed and marginalized $n$, so taking the leading order will
provide an upper estimation for the expected overlapping-event number.}

Equation~\eqref{eq:asymptotic marginalizing n} provides a robust estimation of
the overlapping-event number in future GW observations, while simultaneously has
a concise mathematical form. \Referee{For $\lambda \approx 10^5$ and $\epsilon \approx
10^{-3}$, one immediately asserts that there will be up to $100$ overlapping
events per year on average with an uncertainty of about $10$ events.} Moreover,
there is a clear physical meaning for the asymptotic expressions. The
dimensionless parameter, $\epsilon=r{\Delta t}_{\rm th}$, can be interpreted as
an overlapping parameter, representing the expected number of events within the
overlapping threshold. Then, the expected number of overlapping events can be
intuitively understood as the product of the expected number of total events
$\lambda$ and the overlapping parameter $\epsilon$. From this point of view, one
can also formally define a succinct ``overlapping rate'', as $r_{o} = {\rm
E}[S]/\tau = \epsilon r$. 

\Referee{
Next, we compare our results with previous studies. \citet{Pizzati:2021apa} also analytically estimated the expectation of the
overlapping-event number. They divided the observation in time chunks of size $\Delta t_{\rm th}$,
and estimated the number of chunks with more than two events, denoted as
$N_{k\geq 2}$ in their work. They found that
${\rm E}\big[N_{k\geq 2}\big] =
\big[1-e^{-\epsilon}(1+\epsilon)\big]\lambda/\epsilon$, which is consistent with
the simulation results in \cite{Samajdar:2021egv}. In the limit of $\epsilon \ll
1$, the expectation and variance of $N_{k\geq 2}$
are\footnote{\cite{Pizzati:2021apa} only gave the expectation and discussed its
asymptotic behavior. Since the numbers of events in different time chunks are
independent, $N_{k\geq 2}$ follows a Binomial distribution and the variance can be
easily calculated. Here we follow their model and give the variance for a more complete comparison.}
\begin{equation*}
	\begin{aligned}
        {\rm E}\big[N_{k\geq 2}\big] = \frac{\lambda\epsilon}{2}+{\cal O}(\lambda\epsilon^2)\,,\ \ 
        {\rm Var}\big[N_{k\geq 2}\big] = \frac{\lambda\epsilon}{2}+{\cal
        O}(\lambda\epsilon^2)\,,
    \end{aligned}
\end{equation*}
both differing from Eq.~\eqref{eq:asymptotic marginalizing n} by a factor of
$2$. In our definition, $S$ counts the number of closing pairs of events.  Since
a closing pair has the equal probability of falling in the same chunk or in
adjacent chunks, the expected number $N_{k\geq 2}$ is half of $S$.  It is worth
noting that the chunks are artificial divisions of the observation duration, and
when it comes to data analysis, the closing pairs in adjacent chunks have no
difference with those in the same chunk---both can mislead the search and
parameter estimation.  Therefore, Eq.~\eqref{eq:asymptotic marginalizing n},
along with the exact expression~\eqref{eq:expectation and variance combined with
N}, represents a correct estimation for the overlapping-event number in the view
of data analysis. For the source-dependent threshold model, such as the one in
\citet{Relton:2021cax}, the asymptotic behavior and comparison are described in
Section~\ref{sec:extensions}.}

\subsection{Binomial and Poission Approximations}\label{sec:binomial and poisson
approximations}

In the previous subsection, we find that the asymptotic expectation and variance of
$S$ are equal to each other at the leading order, whether $n$ is fixed or
marginalized. This property reminds us of the Poisson distribution, which has the same
expectation and variance. In addition, $S$ is the sum of $n-1$ identically distributed but
dependent Bernoulli variables with the success
probability $p = {\rm Pr} (I_i = 1) = 1-(1-\xi)^n$, which motivates us to
 consider the relation
between $S$ and a binomial variable $S_{\rm b}\sim {\cal B}(n-1,p)$. In this section,
we
discuss approximating the exact expressions in Section~\ref{sec:exact expressions}
with
more familiar binomial and Poisson distributions.

\begin{figure}[t]
    \centering
    \includegraphics[width=20pc]{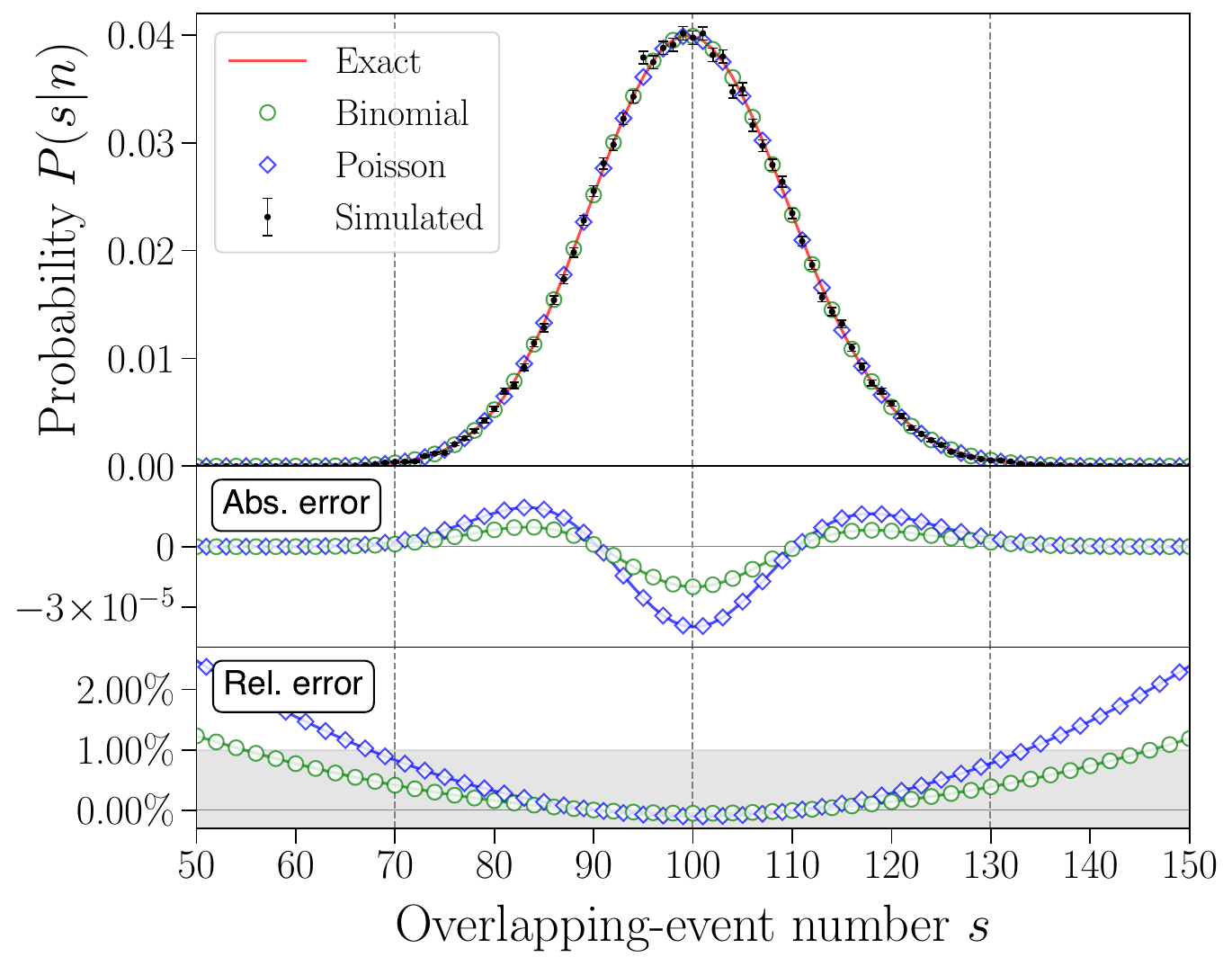}
    \caption{The probability mass function of the overlapping-event number $S$
    for $n=10^5$ and $\epsilon_n = 10^{-3}$. In the upper panel, the red line is
    from the exact expression~\eqref{eq:prob of S explicit}. To avoid
    overcrowding, the binomial (green circle) and Poisson (blue diamond)
    approximation results are shown only for even and odd $s$, respectively. The
    simulation results are shown as black dots with error bars.  The vertical
    dashed lines give the expectation and $3$-$\sigma$ interval of $S$. In the
    two lower panels, we show the absolute and relative errors of the binomial
    and Poisson approximations. The gray shade highlights area where the
    absolute value of relative error is below $1\%$.}
    \label{fig1:overlap number for different distribution}
\end{figure}
For the probability mass function, in Fig.~\ref{fig1:overlap number for
different distribution}, we present the exact
expression~\eqref{eq:prob of S explicit}, and two approximations, ${\cal
B}(n-1,p)$ and ${\rm Pois}\big((n-1)p\big)$ with $n=10^5$ and
$\epsilon_n = 10^{-3}$. Since ${\rm Pois}\big((n-1)p\big)$
is a further approximation of ${\cal B}(n-1,p)$, it has a larger error than the
binomial one. Both
approximations give larger (conservative) variances than the exact
result, so they are more likely to underestimate the probability around the
expectation $s\approx 100$ and overestimate the probability far away from the
expectation. 
Therefore, the absolute errors exhibit two zero points at $s\approx 90$
and $110$ where the errors change sign.  The absolute errors are ${\cal O}
\big(10^{-5}\big)$, while the relative errors are smaller than ${\cal O}
\big(10^{-2}\big)$ within the $3$-$\sigma$ region ($70 \lesssim s \lesssim
130$). We also show the frequency of $S$ in $10^5$ simulations, and the results
are consistent with the analytical expression. 
\Referee{Then we discuss the approximation for the expectation and variance of $S$ given $n$. Due to the
additive property of expectation, the binomial expectation ${\rm
E}[S_{\rm b}|n]$ is same as ${\rm E}[S|n]$. For the variance, we have the
relation
\begin{equation}
    {\rm Var}[S|n] = {\rm Var}[S_{\rm b}|n] + (n-1)(n-2){\rm Cov}(I_i,I_j)\,,
\end{equation}
where the covariance term is negative and corresponds to a ${\cal
O}(\epsilon_n)$ relative difference between ${\rm Var}[S|n]$ and ${\rm
Var}[S_{\rm b}|n]$. It is worth emphasizing again that the negative
covariance term is particularly useful in estimating the variance of $S$,
as it allows a conservative estimation from the well-known variance of the binomial distribution, ${\rm Var}[S_{\rm b}|n] = (n-1)p(1-p)$.
Furthermore, for a large $n$ and a small $p$, the binomial distribution
approaches the Poisson distribution, $S_{\rm P}\sim{\rm Pois}\big((n-1)p\big)$, which has the same
expectation as $S_{\rm b}$ and a slightly larger variance $(n-1)p$. Therefore, the Poisson
approximation also provides the same expectation of $S$ and a
conservative estimation for ${\rm Var}[S|n]$. Further calculations show that the
variance difference between $S$, $S_{\rm b}$ and $S_{\rm P}$ is ${\cal
O}(\epsilon_n)$ smaller than the variance itself.
}

In the marginalized-$n$ case, the expectation and variance from
both approximation becomes inconvenient
to use. For the expectation, both
approximations give the same result as the exact expression in
Eq.~\eqref{eq:expectation of S}, so after marginalizing $n$, the
expectation is still the same as the lengthy expression in Eq.~\eqref{eq:expectation and variance combined
with N}.
For the variance, though it is still possible to analytically calculate variance
of $S_{\rm b}$ and $S_{\rm P}$ after
marginalizing with the Poisson variable $N$, the expressions are still
complicated, showing no advantage over the exact expressions in
Eq.~\eqref{eq:expectation and variance combined with N}. Furthermore, we find
that the relative errors of variance from both approximations are also at the
order of ${\cal O}(\epsilon)$. 
Therefore, in the marginalized-$n$ case we recommend using the leading-order terms in
Eqs.~\eqref{eq:asymptotic marginalizing n} as a quick estimation, whose 
relative errors are also at the order of ${\cal O}(\epsilon)$.

\section{Extensions to the Overlapping-Event Model}\label{sec:extensions}
In previous sections, we rigorously calculated the overlapping-event number $S$
in future GW observations, including the exact distribution, expectation and
variance, as well as its asymptotic expressions and approximations. In the
derivations, the definition of $S$ is relatively simple, only counting the
two-signal overlapping events with a fixed threshold ${\Delta t}_{\rm th}$ and
ignoring the overlaps between different types of sources. This definition is
simple and intuitive, but there also exists more complicated definitions 
in literature for better characterizing 
the overlapping-event population~\citep{Pizzati:2021apa,
Relton:2021cax, Hu:2022bji}. In this section, we extend the model in
Section~\ref{sec:assumptions} to more sophisticated definitions, while still
keeping its analytic nature.

\subsection{Three or More Signal Overlaps}\label{sec:three or more events
overlap}
When the detection rate is high, or the threshold is large, it is possible to have
three or more signals overlapping with each other~\citep{Regimbau:2012ir,
Wu:2022pyg}. Though the probability of three or more
overlapping events is expected to be smaller, if happened, they cause new problems in the data analysis. Currently, the number of such events is usually obtained as the byproduct of
simulating the two-signal overlapping events~\citep{Hu:2022bji}, lacking
a more quantitative estimation. Our model in
Section~\ref{sec:assumptions} can be easily extended to three or more signals,
and here we analytically calculate this kind of
overlapping events for the first time.

As an example, we consider the three-signal overlapping number. Still choosing
$\xi$ as the threshold, we define $S^{(3)} = \sum_{i=3}^n {\cal
I}_{[0,\xi]}\big(\Delta X^{(3)}_i\big)$, with the indicator function ${\cal
I}$ and $\Delta X^{(3)}_i$,
\begin{equation}
    \Delta X^{(3)}_i  =   
        \Delta X_i+\Delta X_{i-1}, \ \  2\leq i\leq n\,.
\end{equation}
For simplicity, we only calculate the expectation of $S^{(3)}$ given $n$ events. Since the joint distributions of $\Delta X_i$ and $\Delta X_{i-1}$ are the
same for all $i$, and the one-dimensional marginal distributions of all
$X^{(3)}_i$ are the same. Taking $i=2$, we find that $\Delta X^{(3)}_2 = X_2$, whose
distribution is the ${\rm Beta}(2,n-1)$ distribution (see 
Appendix~\ref{appsub:joint and marginal distributions of time differences}).
Similar to the expectation of $S$ in
Eq.~\eqref{eq:expectation of S}, the expectation of
$S^{(3)}$ reads
\begin{equation}
    \begin{aligned}
        {\rm E}[S^{(3)}|n]&=
        (n-2)\big[1-(1-\xi)^{n-1}(n\xi-\xi+1)\big]\\
        & = \frac{(n-1)(n-2)}{2n}\epsilon_n^2 + {\cal O}(n\epsilon_n^3) \,,
    \end{aligned}
\label{eq:expectation of S3}
\end{equation}
where the second line shows the leading term for small $\epsilon_n$. Comparing
this with Eq.~\eqref{eq:asymptotic fix n}, we find that
 ${\rm E}[S^{(3)}|n]$ is ${\cal O}(\epsilon_n)$ smaller than the
two-signal overlapping expectation ${\rm E}[S|n]$ at their leading orders.
This is consistent with our intuition that three-signal overlapping
is less likely to occur, and its number is roughly reduced by $\epsilon_n$.

For more than three-signal overlapping events, it can be found that the time
difference between the $i$-th and $(i-m+1)$-th events, denoted as $\Delta X^{(m)}_i$, follows the ${\rm
Beta}(m-1,n-m+2)$ distribution for all $m-1\leq i\leq n$. Similar to the 
three-signal overlapping case, 
the
expectation of the $m$-signal overlapping number ${\rm E}[S^{(m)}|n]$ can be
expressed as
\begin{align}
        {\rm E}[S^{(m)}|n]  &= (n-m+1){\rm Pr}\Big\{\big(\Delta X^{(m)}_i \leq \xi\big)\Big\} \notag\\
        &= \frac{n!}{(m-2)!(n-m)!}\int_0^{\xi} x^{m-2}(1-x)^{n-m+1}{\rm d}x  \,,
        \label{eq:expectation of Sm}
\end{align}
where the integral can be calculated in an explicit but lengthy form. Therefore, it
may be more useful to give a simple upper bound in practice, that is ${\rm E}[S^{(m)}|n] < {n\epsilon_n^{m-1}}/{[(m-1)!]}$.
It can be shown that the leading term of ${\rm E}[S^{(m)}|n]$ is also
at the same order as the upper bound in Eq.~\eqref{eq:expectation of Sm},
from which we find that the expectation of $S^{(m)}$ is less than
${\cal O}\big(\epsilon_n^{m-1}\big)$.

We now substitute the realistic values of $\epsilon_n$ and $n$ to estimate the
overlapping-event number with three or more signals. As a relaxed criterion, the
time difference threshold can be set to be seconds~\citep{Samajdar:2021egv, Hu:2022bji}, corresponding
to $\xi = 10^{-7}$ for a one-year observation duration. For the optimistic
detection rate, we take $n = 10^5$ for BBH events, and $n = 10^6$ for BNS
events, which further gives $\epsilon_n = 10^{-2}$ and $10^{-1}$, respectively for BBHs and BNSs.
According to Eq.~\eqref{eq:expectation of S3}, we expect that there will be less
than $5$ three-signal overlapping events for BBH and about $5000$ events for
BNS events. In other words, the three-signal overlapping events are quite rare for
BBHs (less than $0.005\%$), while there is a non-negligible probability of
three-signal overlapping events for BNSs (about $0.5\%$). Furthermore, there can also be tens
of four-signal overlapping events and several five-signal overlapping
events for BNSs according to Eq.~\eqref{eq:expectation of Sm}. 
On the other hand, if one
wants to safely
ignore the overlapping events involving three signals---less than 10 times per year, for instance---the threshold would need to be under $0.1\,{\rm
s}$ for a
detection rate of $10^6$. However, current studies suggest that this is too
strict~\citep{Samajdar:2021egv, Relton:2021cax, Pizzati:2021apa,
Antonelli:2021vwg}. In fact, with such a high detection rate of BNSs, multiple-signal overlapping events are
inevitable, and the data analysis is required to deal with multiple events merging within
a few seconds. There are some studies that began to develop algorithms to tackle
this challenge~\citep{Miller:2023rnn, Hu:2025vlp}.

\subsection{Overlaps between Different Types of Sources}\label{sec:overlaps
between different types of sources}
Here we extend the model to consider the overlaps between different types of
sources, such as the overlaps between BNS and BBH events. In this case, we
denote the detection rate of BNSs and BBHs as $r_1$ and $r_2$, respectively,
and the total detection rate is $r = r_1 + r_2$. For the overlap variable, we
now use $I_i^{1-1}$, $I_i^{1-2}$ and $I_i^{2-2}$ to denote the closing pairs
with BNS-BNS, BNS-BBH/BBH-BNS and BBH-BBH types of events, respectively. 
Since the types of
events are independent with their arrival times, the expectation of the
indicator variables can be calculated as
\begin{equation}
    \begin{aligned}
            {\rm E}[I_i^{1-1}|n] &= \frac{r_1^2}{r^2}{\rm E}[I_i|n] \,,\\
            {\rm E}[I_i^{1-2}|n] &= \frac{2r_1 r_2}{r^2}{\rm E}[I_i|n] \,,\\
            {\rm E}[I_i^{2-2}|n] &= \frac{r_2^2}{r^2}{\rm E}[I_i|n] \,,
    \end{aligned}\label{eq:expectation of I for different types}
\end{equation}
where ${\rm E}[I_i|n]$ is the expectation of the indicator variable for all types of
events, given in Eq.~\eqref{eq:expectation of S}. Then, the number of overlapping
events with different types of sources, denoted as $S^{1-1}$, $S^{1-2}$ and
$S^{2-2}$, are the addition of corresponding indicator
variables, whose expectations are simply the expectation of all types of events multiplied by the
corresponding coefficients in Eq.~\eqref{eq:expectation of I for different
types}. Furthermore, the types of events are independent with the total event
number, so after marginalizing $n$, the expectations of $S^{1-1}$, $S^{1-2}$ and
$S^{2-2}$ are still the multiplication of ${\rm E}[S]$ in Eq.~\eqref{eq:expectation of S} and the
coefficients in Eq.~\eqref{eq:expectation of I for different types}.

Similar to the previous subsection, we numerically calculate these expectations.
We take the optimistic detection rates of $r_1 = 10^6\,{\rm yr}^{-1}$ and
$r_2 = 10^5\,{\rm yr}^{-1}$, and a relaxed threshold of $\xi = 10^{-7}$. With
these choices, for a one-year observation, we have ${\rm E}[S^{1-1}] \approx 9.5\times 10^4$,
${\rm E}[S^{2-2}] \approx 9.5\times 10^2$ and ${\rm E}[S^{1-2}] \approx 1.9\times 10^4$. For
the asymptotic expressions, we only need to multiply the coefficients in
Eq.~\eqref{eq:expectation of I for different types} with the leading order of
${\rm E}[S]$ in Eq.~\eqref{eq:asymptotic marginalizing n}. For the
BNS-BBH/BBH-BNS overlapping events, at the leading order we have
\begin{equation}
    {\rm E}[S^{1-2}] \approx \frac{2r_1 r_2}{r^2}\lambda\epsilon  = 2\lambda_1\lambda_2\xi\,,
\end{equation}
where $\lambda_1 = r_1\tau$ and $\lambda_2 = r_2\tau$ are the expected number of
BNS and BBH events in the observation duration $\tau$, respectively.
Substituting the realistic values, the leading order gives $2\times 10^4$ for
${\rm E}[S^{1-2}]$, which is consistent with the exact result (at a relative error
of ${\cal O}(\epsilon)$). For the BNS-BNS overlapping events, at the leading
order we have 
\begin{equation}
    {\rm E}[S^{1-1}] \approx \frac{r_1^2}{r^2}\lambda\epsilon = \lambda_1\epsilon_1 \,,
\end{equation}
with $\epsilon_1 = r_1{\Delta t}_{\rm th}$, which reduces to the case of only
considering one type of sources.
 Similar conclusion also holds for the
BBH-BBH overlapping events, where the expectation is ${\rm E}[S^{2-2}] \approx
\lambda_2\epsilon_2$ with $\epsilon_2 = r_2{\Delta t}_{\rm th}$. 
\cite{Pizzati:2021apa} also calculated the overlapping events between
different types of sources and obtained an expectation of $\lambda_1\lambda_2\xi$
for the BNS-BBH/BBH-BNS overlapping events, where we observe once again that the
difference in a factor of $2$ as discussed in Section~\ref{sec:asymptotic
expressions}. In addition, directly
multiplying the coefficients in Eq.~\eqref{eq:expectation of I for different
types} with the exact expression in Eq.~\eqref{eq:expectation and variance
combined with N} does not give the same result of only considering one type of
sources. Consider a three-signal overlapping case where one BBH event is in between two BNS
events, and the time difference between the two BNS events is smaller than
${\Delta t}_{\rm th}$. The two BNS events will contribute to $S^{1-2}$ instead of
$S^{1-1}$.
Since this only happens in three or more signal overlapping events, the difference is at
the next-to-leading order, and ${\rm E}[S^{1-1}]$ equals to ${\rm E}[S]$ (only
considering BNS events) at the leading order.

Here we only calculate the expectation as
an illustration. But it is also possible to analytically calculate the corresponding
variance of $S^{1-2}$ by rederiving 
the covariance between the indicator
variables, ${\rm Cov}\big(I_i^{1-2},I_j^{1-2}\big)$. In addition, this framework
can be easily extended to cases considering the overlaps between more
types of sources, such as neutron star--black hole (NSBH) events, where the coefficients in
Eq.~\eqref{eq:expectation of I for different types} should be replaced with the fraction of considered types of sources.

\subsection{Source-Dependent Thresholds}\label{sec:source-dependent thresholds}

In the discussion above, we have assumed a fixed threshold ${\Delta t}_{\rm
th}$ for determining the overlapping events, which is a choice made for
simplicity. However, it has been shown that the overlapping signals have
significant effects on the data analysis only when the time-frequency tracks of
the individual signals are crossing~\citep{Wang:2023ldq, Johnson:2024foj}. Since
sources with different parameters have different frequency evolutions,
considering the source-dependent thresholds serves as a better assessment of the
overlapping events in the view of data analysis. 

To extend our model to the source-dependent thresholds, we first modify the
definition of the indicator variable: $I_i = {\cal I}_{[0,\Delta t_{\rm
th}(\Theta_{i-1},\Theta_{i})]}\big(\Delta T_i\big)$, where $\Delta t_{\rm
th}(\Theta_1,\Theta_2)$ now depends on the parameters of the two adjacent
sources, $\Theta_1$ and $\Theta_2$. For example, the threshold can be defined by requiring the two
signals have a time-frequency track crossing and the crossing frequency is
within the sensitive band of the detector.
 We also assume that the parameters of
observed sources are independently drawn from the observed distribution
$\pi$.
 The expectation of $S$ now reads
    \begin{align}
        {\rm E}[S|n] &= \sum_{i=2}^n \int {\rm E}\big[I_i|n,\theta_{i-1},\theta_i\big] \pi(\theta_{i-1}) \pi(\theta_i) {\rm d}\theta_{i-1} {\rm d}\theta_i\notag \\
        &= (n-1)\Big \langle 1 - \big(1-\xi(\theta_{1},\theta_2)\big)^n \Big\rangle \,,\label{eq:expectation of S source-dependent}
    \end{align}
where $\xi(\theta_1,\theta_2)$ is the dimensionless source-dependent threshold,
and in the second line we have used the independence between the source parameters
and their arrival times. We also introduce $\langle \cdot
\rangle$ to denote the average over the parameters of the
two adjacent sources. Since the total event number is also independent of the
source parameters, the expectation of $S$ after marginalizing $n$ is
\begin{equation}
    {\rm E}[S] = \Big\langle (\lambda - 1)(1 - e^{-\epsilon}) + \epsilon
    e^{-\epsilon}\Big\rangle \,,\label{eq:expectation of S source-dependent marginalizing n}
\end{equation}
where $\epsilon(\theta_1,\theta_2) =  \lambda\xi(\theta_1,\theta_2)$ is a
function of the source parameters. 
Similar to their counterparts in the constant-threshold case, it is useful to find the
asymptotic expressions of Eqs.~\eqref{eq:expectation of S source-dependent} and
\eqref{eq:expectation of S source-dependent marginalizing n}. Here we only
discuss the latter one for brevity. Assuming that $\epsilon(\theta_1,\theta_2)$ has a
supreme $\epsilon_{\rm max}$, in the limit of $\epsilon_{\rm max} \ll 1$, the
leading order of Eq.~\eqref{eq:expectation of S source-dependent marginalizing
n} is
\begin{equation}
    {\rm E}[S] \approx \lambda\langle \epsilon
    \rangle  = \lambda \int \epsilon(\theta_1,\theta_2) \pi(\theta_1) \pi(\theta_2)
{\rm d}\theta_1 {\rm d}\theta_2\,,
\label{eq:expectation of S source-dependent marginalizing n asymptotic}
\end{equation}
where $\langle \epsilon \rangle$ has a clear physical meaning, the average of the overlapping
parameter over the observed population. Equation~\eqref{eq:expectation of S source-dependent marginalizing n
asymptotic} can be regarded as the extension of the asymptotic
expectation in Eq.~\eqref{eq:asymptotic marginalizing n} to the source-dependent
thresholds, while still preserving its concise form and intuitive interpretation.

The overlapping number with source-dependent thresholds has been
analytically explored in \citet{Relton:2021cax}, where the authors modelled the
time differences between the adjacent sources as i.i.d.\ exponential variables.
However, when considering the source dependence, they only
averaged the overlapping parameter $\epsilon$ in the exponential part,
instead of averaging the whole conditional expectations like in Eqs.~\eqref{eq:expectation of S source-dependent} and
\eqref{eq:expectation of S source-dependent marginalizing n}. As a result, 
their results deviate from the exact results from the second
order of $\epsilon$.
 In addition,
the threshold in \citet{Relton:2021cax} was chosen to be the effective duration
of the second signal in the pair, and the average is only over one set of source
parameters. Since the crossing of time-frequency tracks depends on both signals, a more general and realistic threshold should depend on both sets of
source parameters. We verified that, in the case that the threshold only depends on the second signal, the expectation of $S$ in Eq.~\eqref{eq:expectation of S source-dependent marginalizing n} agrees with the
result in \citet{Relton:2021cax} at the leading order of $\epsilon$.

\section{Distribution of Time Differences in an Observation Run}\label{sec:distribution of time differences}

When only counting the overlapping-event number $S$, we lose information about
the merging time difference. As shown in Appendix~\ref{appsub:joint and marginal
distributions of time differences}, the marginal distribution of the
dimensionless time
difference ${\Delta X}_i$ is the beta distribution ${\rm Beta}(1,n)$,
representing the probability density in repeated runs. However, practically
one may also
interest in the distribution of time differences in a single observation run.
More specifically, suppose that there are $n$ events in an observation duration
$\tau$, one has $n-1$ successive time differences, and can obtain a
``distribution'' of ${\Delta T}$ by plotting them in a histogram. Strictly
speaking, the histogram does not represent a distribution function, since the
``distribution'' is different in each observation run, and essentially it is a
random field. In this work, we simply call the histogram a ``distribution'' and
use ``in a single observation'' to emphasize its randomness.
It displays the number of events with ${\Delta T}$ in the histogram bins, serving as a
more detailed description for the overlapping-event population.

\begin{figure}[t]
    \centering
    \includegraphics[width=20pc]{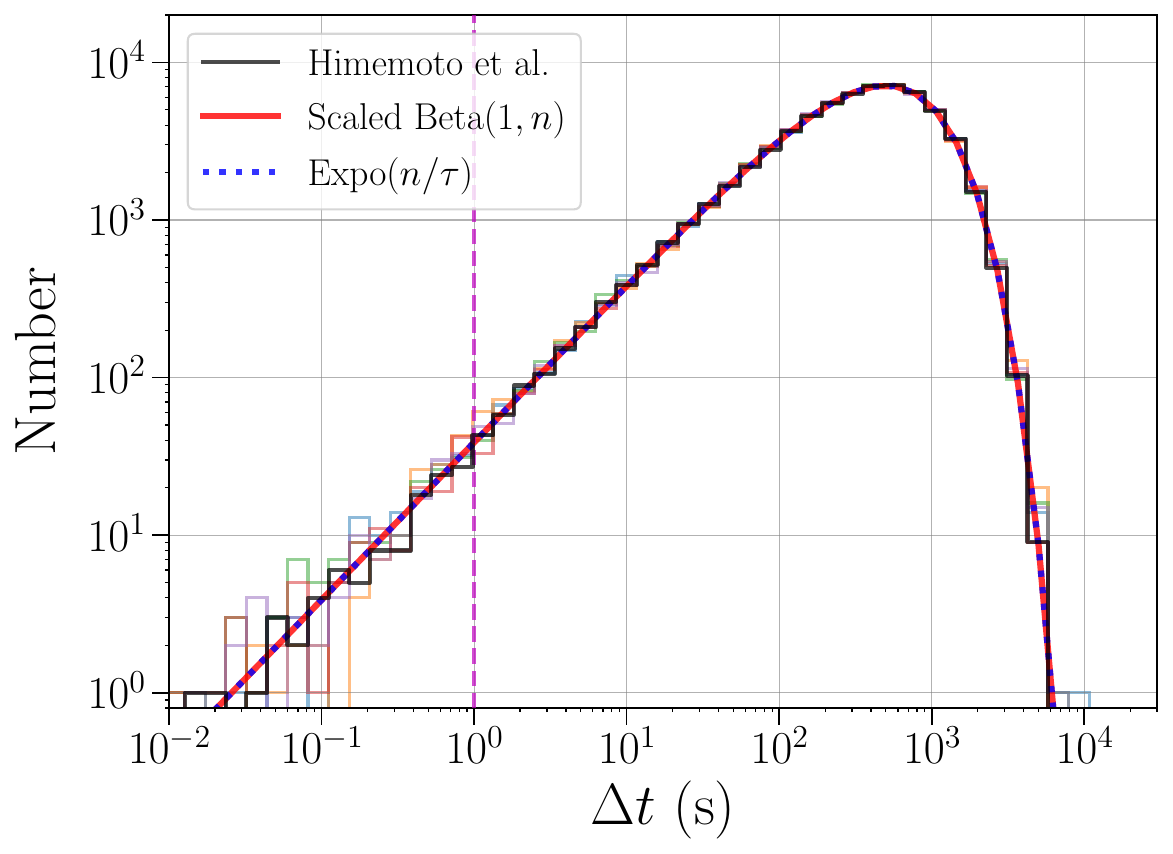}
    \caption{The distribution of time differences in a single observation with
    $n=63100$ and $\tau=1\,{\rm yr}$. The black line shows the result by
    Himemoto et al.~\citep{Himemoto:2021ukb}, while the lines with light colors
    are generated from 5 independent simulations. The expected event numbers
    calculated from the ${\rm Beta}(1,n)$ distribution (scaled by $\tau$) and
    the ${\rm Expo}(n/\tau)$ distribution are plotted as red and blue lines,
    respectively. }
    \label{fig2:compare}
\end{figure}

Recently, \citet{Himemoto:2021ukb} simulated the ${\Delta T}$
distribution in a single observation for BBH and BNS events. In
Fig.~\ref{fig2:compare}, we show their results for BNS events in black, where
the observation duration $\tau$ is $1\,{\rm yr}$ with $ n = 63100$ events. With
the same event number, we also simulate the ${\Delta T}$ distribution for 5
times, plotted in light colors. Except for some statistical fluctuations, the
distributions agree with each other, even though they were generated from
different and independent simulations. Furthermore, these distributions are
consistent with the shape of the ${\rm Beta}(1,n)$ distribution (scaled by
$\tau$), plotted as a red line. 

In fact, the expected distribution of ${\Delta X}$ in a single observation is
exactly the ${\rm Beta}(1,n)$ distribution. Consider that the expected number of
time differences below $\xi$ is exactly ${\rm E}[S|n]$ in
Eq.~\eqref{eq:expectation of S}, the expected distribution of ${\Delta X}$ can
be derived as
\begin{align}
    P({{\Delta x}}|n) &= \frac{1}{n-1}{\rm E}\left(\!\left .\! \frac{{\rm d}
    \,{\rm Num}(\Delta X \leq \xi)}{{\rm d}\xi}\right |_{\xi= {{\Delta x}}}
    \!\right) \notag \\
    &= n\left(1-{{\Delta x}}\right)^{n-1}\,,
    \label{eq7:expected distribution of time differences}
\end{align}
which is the ${\rm Beta}(1,n)$ distribution. Similar to the derivation
of the expectation and variance of $S$, in Appendix~\ref{appsub:expectation and
variance of S} we find that the number of differences
in a dimensionless histogram bin $[\xi,\xi+{\rm  d}\xi]$, denoted as ${\rm  d}
S$, has the expectation ${\rm E}[{\rm  d} S|n] = (n-1)\cdot n(1-\xi)^{n-1}{\rm
d} \xi$ at the leading order of ${\rm  d} \xi$. The variance of ${\rm  d}S$ also
equals to the expectation ${\rm E}[{\rm  d} S|n]$ at the leading order. It
suggests that the simulated distribution aligns better with the ${\rm
Beta}(1,n)$ distribution in bins where ${{\rm E}[{\rm  d} S|n]} \gg 1$, while
the fluctuations are larger in bins where ${{\rm E}[{\rm  d} S|n]} \sim 1$, as
seen in Fig.~\ref{fig2:compare}.

\Referee{It is also interesting to discuss the ${\Delta X}$ distribution when considering overlaps between
different types of sources. For example, if we inject $n_1$
BNS signals and $n_2$ BBH signals, the expected distribution of ${\Delta X}$
becomes ${\rm Beta}(1,n_1+n_2)$ since there are $n_1+n_2$ events in total.
Because of the independence of the arrival times and the event types, the
expected distribution of BNS-BBH/BBH-BNS overlaps is also ${\rm
Beta}(1,n_1+n_2)$. Furthermore, the distribution for BNS-BNS or BBH-BBH overlaps
is also ${\rm Beta}(1,n_1+n_2)$, instead of ${\rm Beta}(1,n_1)$ or ${\rm
Beta}(1,n_2)$ in the case of only considering one type of sources. 
Adding new types of overlapping sources is more likely to separate 
adjacent events with larger separation, which
changes the distribution of time differences among all sources.}

Finally, we return to the relation between our model and the Poisson process. In
the Poisson process, the time differences between adjacent events are
independently distributed as the exponential distribution ${\rm Expo}(r)$. There
are some investigations using the exponential distribution ${\rm Expo}(r)$ to
model the merger time differences and calculate the overlapping-event
rates~\citep{Relton:2021cax}. In fact, the scaled ${\rm Beta}(1,n)$ distribution
approaches the exponential distribution ${\rm Expo}(n/\tau)$ for large $n$ and
$\tau$,
\begin{equation}
    P(\Delta t) = \frac{n}{\tau}\left(1-\frac{\Delta
    t}{\tau}\right)^{n-1}\approx \frac{n}{\tau}e^{-n\Delta t/\tau}\,,
    \label{eq8:beta to expo}
\end{equation}
which is also shown in Fig.~\ref{fig2:compare}. In the long duration limit, we
have $n/\tau \approx r$, and the exponential distribution ${\rm Expo}(r)$ in the
Poisson process is recovered as expected. This explains the Poisson-like
behaviors of the expectation and variance of $S$ in our model at the leading
order. However, it should be emphasized that the time differences are
identically distributed but not independent. In Appendix~\ref{appsub:relation to Poisson process}, we prove
that the joint distribution of the time differences can be regarded as the
conditional distribution of the first $n$ time differences in the Poisson
process under the condition of observing $n$ events in a finite duration $\tau$. In realistic observations, the translation symmetry in a
Poisson process is broken, and it is more appropriate to use the joint
distribution of the time differences in Eq.~\eqref{eq:joint distribution of dimensionless time difference} to account for the
correlations between them. 

\section{Conclusions and Discussions}\label{sec:conclusions}

In this work, we rigorously derived the distribution of the overlapping-event number $S$ for a
given GW event number $n$. The formulae for the expectation and variance of $S$
depend on two dimensionless parameters, $\lambda = r\tau$ and $\epsilon =
r{\Delta t}_{\rm th}$. We discussed the validity of binomial and Poission
approximations in the context of XG GW detectors. At the leading order, the
overlapping-event number is $\lambda\epsilon\pm \sqrt{\lambda\epsilon}$ with
negative corrections at ${\cal O}\big(\epsilon\big)$.  
We conduct analytical and quantitative discussions for
the distribution of the overlapping-event number in XG GW observations, providing a
rigorous theoretical foundation for further studies.

\Referee{We also extend the analytical framework to more sophisticated
scenarios, such as multiple-signal overlapping events, overlaps between
different types of sources, and source-dependent thresholds. For the
multiple-signal overlapping events, we firstly derive the expectation of the
three-signal overlapping number, and established an upper bound for the
expectation of overlapping events consisting of more than three signals.  For
the overlaps between different types of sources, we find our results differing
from \citet{Pizzati:2021apa} by a factor of 2, same as in the case of only
considering one type of sources. This is because the definition in
\citet{Pizzati:2021apa} counts the number of chunks with more than two events,
ignoring the overlapping signals spanning two chunks. Apart from this, however,
our results are consistent with the simulation results in
\citet{Samajdar:2021egv} and the analytical expression in
\citet{Pizzati:2021apa}.  For the source-dependent thresholds, we rigorously
derive the expectation of the overlapping-event number, and find that the
expression is an average over the whole conditional expectation for the two
adjacent sources.  In \citet{Relton:2021cax}, the average was only taken over
the exponential part and only for one source parameter, which leads to a
deviation at the next-to-leading order.  We leave more detailed discussions that
substitute specific source-dependent threshold models and population models to
future works.
}

We also proved that the distribution of time differences in a single observation is
the beta distribution on average, which analytically explains the simulation
results in previous studies~\citep{Himemoto:2021ukb}. In our work, the merger
time differences are modeled as the differences between $n$ i.i.d.\ uniform
variables. This distribution can be regarded as the conditional distribution of
the first $n-1$ inter-arrival times in a Poisson process within a finite
observation duration, which is a more realistic model for GW observations. It is
interesting to see how this
distribution serves as a useful prior reference for future search and parameter
estimation of overlapping signals.
\begin{acknowledgments}
We thank Yiming Dong and Zexuan Wu for helpful discussions, \Referee{and the anonymous referee for comments}.  
This work was supported by the National Natural Science Foundation of China
(123B2043), the Beijing Natural Science Foundation (1242018), the National SKA
Program of China (2020SKA0120300), the Max Planck Partner Group Program funded
by the Max Planck Society, and the High-performance Computing Platform of Peking
University.  Z.H.\ is supported by the China Scholarship Council (CSC).

\end{acknowledgments}

\appendix

\section{Derivation of Formulae in the Main Text}\label{sec:appendix}
\subsection{Joint and Marginal Distributions of Time Differences}\label{appsub:joint and marginal distributions of time differences}
\Referee{Here we review the joint and marginal distributions of order statistics of $n$ i.i.d.\ random
variables and discuss their useful properties. Similar derivations can be found in
many statistics textbooks, such as~\citet{casella2024statistical}.}Denoting
$f(x)$ as the probability density function of the random variable before
ordering, the joint distribution of their order statistics 
is
\begin{equation}
    \begin{aligned}
        P({\bm x}) = n!\prod_{i=1}^n f(x_i)\,,\ \ 0\leq x_1\leq x_2\leq \cdots
        \leq x_n\leq 1\,.
    \end{aligned}
\end{equation}
For the uniform distribution in this work, $f(x) = 1$. To marginalize some order variables, the
following formula is useful,
\begin{equation}
    \int\limits_{x_a < x_1 < \cdots < x_m < x_b} \prod_{i=1}^m f(x_i) {\rm d}x_i
    = \frac{\big(F(x_b)-F(x_a)\big)^m}{m!}\,,
\end{equation}
where $x_a$ and $x_b$ are two arbitrary points in the support set of $X_i$.
$F(x)$ is the cumulative distribution function of $f(x)$, and for uniform
distribution we have $F(x)
= x$. For example, the one-dimensional marginal distribution of $X_i$ is
\begin{equation}
    \begin{aligned}
        P(x_i)& =  n! f(x_i)\int\limits_{0<x_1<\cdots<x_{i-1}<x_i}
        \prod_{k=1}^{i-1} f(x_k) {\rm d}x^k
        \int\limits_{x_i<x_{i+1}<\cdots<x_n<1} \prod_{k=i+1}^{n} f(x_k) {\rm
        d}x^k\\ 
        &= f(x_i)\frac{n!F(x_i)^{i-1}\big(1-F(x_i)\big)^{n-i}}{(i-1)!(n-i)!} 
        = \frac{n!}{(i-1)!(n-i)!}x_i^{i-1}\big(1-x_i\big)^{n-i}\,,
    \end{aligned}\label{eqs:marginal distribution of time difference}
\end{equation}
which is the ${\rm Beta}(i,n-i+1)$ distribution.

The differences between adjacent order statistics are defined as 
\begin{equation}
    \Delta X_i =
    \begin{cases}
        X_i, & i = 1 \\
        X_i - X_{i-1}, & i = 2,3,\cdots,n
    \end{cases}\,,
\end{equation}
which can be regarded as a variable transformation from ${\bm x}$ to $\Delta
{\bm x}$ with the Jacobian
\begin{equation}
\left|\frac{\partial \Delta{\bm X}}{\partial {\bm X}}\right| =
    \begin{vmatrix}
        1 & 0  & 0  & \cdots & 0  \\
        -1 & 1  & 0  & \cdots & 0  \\
        0  & -1 & 1  & \cdots & 0  \\
        \vdots & \vdots & \vdots & \ddots & \vdots \\
        0  & 0  & 0  & \cdots & 1
    \end{vmatrix} = 1\,.
\end{equation}
Therefore, the joint distribution of $\Delta {\bm X}$ reads
\begin{equation}
    P(\Delta {\bm  x}) = n!\,,\ \  {{\Delta x}}_i\geq 0\ \  \& \ \  {\sum
    \nolimits}_{i=1}^n {{\Delta x}}_i\leq 1\,.
\end{equation}
Note the difference of support sets due to the transformation. This is
Eq.~\eqref{eq:joint distribution of dimensionless time difference} in the main
text.  This
distribution can be understood as a uniform distribution within the region
enclosed by the $n-1$ dimensional standard simplex and the coordinate planes,
and its symmetry about each component is crucial for our calculations. As a more mathematical statement, the joint distribution of $\Delta
X_1,\Delta X_2,\cdots,\Delta X_{n}$ and $1 - X_n$ is the symmetric Dirichlet
distribution with parameter $\alpha =1$. 

To calculate the marginal distributions, such as the joint distribution of two
differences, $\Delta X_i$ and $\Delta X_j$, a common approach is to first
calculate the joint distribution of $X_i$ and $X_j$, and then convert it to the
$\Delta X$ space. However, benefiting from the special form of the joint
distribution of $\Delta X$ in our case, we can directly calculate the
probability that $\Delta X_i > \xi_i$ ($i=1, \cdots, m$)
according to the geometry of the simplex, i.e., Eq.~\eqref{eq:prob of
dimensionless time difference} in the main text. Taking the derivative of the above equation
with respect to $\xi_i$, we obtain the joint distribution of $\Delta
X_1,\cdots,\Delta X_m$ as 
\begin{equation}
    P(\Delta x_1,\cdots,\Delta x_m) =
    \frac{n!}{(n-m)!}\left(1-\sum_{i=1}^{m}\xi_i\right)^{n-m}\,.
    \label{eqs:joint distribution of time differences}
\end{equation}
Due to the symmetry of the simplex, this distribution function is the same for
any $m$ components of $\Delta {\bm X}$, and the marginal distribution of $\Delta
X_i$ is the ${\rm Beta}(1,n)$ distribution.  Especially, taking $m=1$, the
marginal distribution of $\Delta X_i$ is 
\begin{equation}
    P({{\Delta x}}_i) = n(1-{{\Delta x}}_i)^{n-1}\,,
\end{equation}
which is the ${\rm Beta}(1,n)$ distribution.

\subsection{Distribution of the Overlapping-Event Number with Known Total Event
Number}\label{appsub:derive details of S distribution}

In the first line of Eq.~\eqref{eq:intersection of A_i}, we write the
probability of occurrence of $s$ overlapping events given $n$ events in an
intersection form involving the event $A_i = \big\{I_i = 1\big\}$ and its
complement $\bar{A_i}$. We first reformulate this complex event as
\begin{equation}
    \begin{aligned}
        {\rm Pr}\,\Big (\big(\cap_{i=2}^{s+1} {A}_i\big)\cap
        \big(\cap_{i=s+2}^{n} \bar{A_i}\big)\Big) 
        &={\rm Pr}\,\Big(\cap_{i=s+2}^{n} \bar{A_i}\Big) - {\rm Pr}\,\Big
        (\overline{\cap_{i=2}^{s+1} A_i}\cap \big(\cap_{i=s+2}^{n}
        \bar{A_i}\big)\Big)\\
        &={\rm Pr}\,\Big(\cap_{i=s+2}^{n} \bar{A_i}\Big) - {\rm Pr}\,\Big
        (\big(\cup_{i=2}^{s+1} \bar{A}_i\big)\cap \big(\cap_{i=s+2}^{n}
        \bar{A_i}\big)\Big)\,.
    \end{aligned}
\end{equation}
The second term can be expressed in the intersections of $\bar{A}_i$ with
 the inclusion-exclusion principle,
\begin{align}
    {\rm Pr}\,\Big (\big(\cup_{i=2}^{s+1} \bar{A}_i\big)\cap
    \big(\cap_{i=s+2}^{n} \bar{A_i}\big)\Big) =
    & \sum_{2\leq i_1\leq s+1} P\big(\bar{A}_{i_1}\cap \big(\cap_{i=s+2}^{n}
    \bar{A_i}\big)\Big) - \sum_{2\leq i_1<i_2\leq s+1} {\rm
    Pr}\,\Big(\bar{A}_{i_1}\cap \bar{A}_{i_2}\cap \big(\cap_{i=s+2}^{n}
    \bar{A_i}\big)\Big)\notag \\
    &+\sum_{2\leq i_1<i_2<i_3\leq s+1} {\rm Pr}\,\Big(\bar{A}_{i_1}\cap
    \bar{A}_{i_2}\cap \bar{A}_{i_3}\cap \big(\cap_{i=s+2}^{n}
    \bar{A_i}\big)\Big) + \cdots \notag  \\
    =& \sum_{k=1}^s\left(\sum_{2\leq i_1<\cdots<i_k\leq s+1} (-1)^{k-1}{\rm
    Pr}\,\Big(\big(\cap_{j=1}^k \bar{A}_{i_j}\big)\cap \big(\cap_{i=s+2}^{n}
    \bar{A_i}\big)\Big)\right)\notag \\
    =& \sum_{k=1}^s C_{s}^k(-1)^{k-1}\Big(1-\big(n-1-(s-k)\big)\xi \Big)^n\,,
\end{align}
where in the last line, $C_{s}^k\equiv $ represents the binomial coefficient and we
have used the fact that
\begin{equation}
    {\rm Pr}\, \big(\cap_{i=1}^{m} \bar{A}_i\big) = \left(1-m\xi\right)^n\,.
\end{equation}
Noting that ${\rm
Pr}\,\big(\cap_{i=s+2}^{n} \bar{A_i}\big) = \big(1-(n-s-1)\xi\big)^n$ can be
formally regarded as the $k=0$ term in the above sum, we write the probability
of $s$ overlapping events as 
\begin{equation}
    \begin{aligned}
        P(s|n) &= C_{n-1}^s {\rm Pr}\left(\big (\cap_{i=2}^{s+1} {A}_i\big)\cap
        \big (\cap_{i=s+2}^{n} \bar{A_i}\big)\right)
        = C_{n-1}^s\sum_{k=0}^s C_{s}^k(-1)^{k}\Big[1-(n+k-s-1)\xi \Big]^n\,.
    \end{aligned}
\end{equation}
This is Eq.~\eqref{eq:prob of S explicit} in the
main text, and we leave a further simplification of the summation to future
work.

\subsection{Marginal Expectation and Variance of the Overlapping-Event Number}
\label{appsub:expectation and variance of S}

For computing the expectation and variance of $S$ given $n$, it is convenient to
use the overlapping variable $I_i$ and the event $A_i$ in the calculation, since
the expectation of the product of $m$ variables $I_i$ exactly
corresponds to the probability of the intersection of the associated events
$A_i$.  The expectation is 
\begin{equation}
    \begin{aligned}
        {\rm E}[S|n] = \sum_{i=2}^{n} {\rm E}[I_i|n]
         &= \sum_{i=2}^{n} \left[1\cdot {\rm Pr}\,(A_i) +0\cdot {\rm
         Pr}\,(\bar{A}_i) \right] 
         = (n-1)\Big[ 1-(1-\xi)^n\Big]\,.
    \end{aligned}\label{eqs:expectation of S given n}
\end{equation}
For the variance, we have
\begin{equation}
    \begin{aligned}
        {\rm Var}[S|n] =  \sum_{i=2}^n {\rm Var}[I_i|n] + 2\sum_{2\leq i<j\leq
        n}{\rm Cov}(I_i,I_j)\,,
    \end{aligned}\label{eqs:variance of S given n}
\end{equation}
where the condition of $n$ in the covariance ${\rm Cov}(I_i,I_j)$ is omitted for
brevity. Due to the symmetry in the joint distribution of $\Delta {\bm X}$, all
the pairwise covariances are the same, and can be calculated as
\begin{equation}
    \begin{aligned}
        {\rm Cov}(I_i,I_j) &=  {\rm E}[I_iI_j|n] - {\rm E}[I_i|n]\cdot {\rm E}[I_j|n] \\
        &= P(A_i\cap A_j) - P(A_i)P(A_j)\\
        &= (1-2\xi)^n - (1-\xi)^{2n}\,.
    \end{aligned}
\end{equation}
This covariance is clearly negative since $(1-\xi)^{2n}= (1+\xi^2-2\xi)^n
>(1-2\xi)^n$ for $\xi\in(0,1/2)$. The variance of $I_i$ is  calculated as ${\rm
Var}[I_i|n] = {\rm E}[I_i^2|n] - {\rm E}[I_i|n]^2 =
(1-\xi)^n\big(1-(1-\xi)^n\big)$. Therefore, the variance of $S$ given $n$ is
\begin{equation}
    {\rm Var}[S|n] = (n-1)\Big[(1-\xi)^n+(n-2)(1-2\xi)^n -
    (n-1)(1-\xi)^{2n}\Big]\,.
\end{equation}

For the calculation of the expectation and variance of ${\rm d}S$ in a
dimensionless histogram bin $[\xi,\xi+{\rm d}\xi]$, we have to return to the
distribution functions. For example, the expectation of ${\rm d}S$ is 
\begin{equation}
    \begin{aligned}
        {\rm E}[{\rm d}S|n] &= \sum_{i=2}^{n} {\rm Pr}\,\Big(\big\{\xi\leq
        \Delta X_i<\xi+{\rm d}\xi\big\}\Big) \approx (n-1)\cdot
        n(1-\xi)^{n-1}{\rm d}\xi\,,
    \end{aligned}
\end{equation}
where the factor $(n-1)$ comes from the fact that all the $n-1$ differences are
identically distributed, while $n(1-\xi)^{n-1}$ comes from the one-dimensional
distribution function of $\Delta X_i$, i.e. the ${\rm Beta}(1,n)$ distribution.
To find ${\rm Var}[{\rm d}S|n]$, we further need the joint distribution of two
differences $\Delta X_i$ and $\Delta X_j$ given in Eq.~\eqref{eqs:joint
distribution of time differences}. This calculation is lengthy but
straightforward, and we find that ${\rm Var}[{\rm d}S|n]$ is the same as the
expectation ${\rm E}[{\rm d}S|n]$ at the leading order of ${\rm d}\xi$.

In the given observation duration $\tau$, the number of events $N$ follows the
Poisson distribution ${\rm Pois}(\lambda)$ with $\lambda = r\tau$,
\begin{equation}
    P(n) = \frac{e^{-\lambda}\lambda^n}{n!}\,.
\end{equation}
The expectation of $S$ is then
\begin{equation}
    {\rm E}[S] = \sum_{n=2}^{\infty} {\rm E}[S|n]P(n) = \sum_{n=2}^{\infty}
    (n-1)\Big[ 1-(1-\xi)^n\Big]\frac{e^{-\lambda}\lambda^n}{n!}\,.
    \label{eqs:expectation of S}
\end{equation}
For the variance, we use the formula
\begin{equation}
    \begin{aligned}
        {\rm Var}[S] = {\rm E}[S^2]-{\rm E}[S]^2
         &= \sum_{n=2}^{\infty} {\rm E}[S^2|n]P(n) -{\rm E}[S]^2\\
         & = \sum_{n=2}^{\infty} \left\{\Big({\rm Var}[S|n] + {\rm
         E}[S|n]^2\Big)P(n) \right\}-{\rm E}[S]^2\,,
    \end{aligned}
\end{equation}
where ${\rm E}[S|n]$, ${\rm Var}[S|n]$ and ${\rm E}[S]$ are given in
Eqs.~\eqref{eqs:expectation of S given n}, \eqref{eqs:variance of S given n}
and \eqref{eqs:expectation of S}, respectively. In the calculation, the term
with the shape of $n(\cdot)^n/n!$ and $n^2(\cdot)^n/n!$ can be explicitly summed
up with the properties of the Poisson distribution. After a lengthy calculation,
the expectation and variance of $S$ are obtained as Eq.~\eqref{eq:expectation
and variance combined with N} in the main text.

\subsection{Relation to the Poisson Process}\label{appsub:relation to Poisson process}

Denoting $\Delta \widetilde{\bm T}$ as the first $n$ time differences in Poisson
process, here we prove that Eq.~\eqref{eq:joint distribution of dimensionless
time difference} can be obtained by imposing the condition
$C\equiv\big\{\text{observing}\,n\,\text{events in}\,\tau\big\}$. According to
the Bayes' theorem, we have
\begin{equation}
    P(\Delta \tilde{\bm t}|C) = \frac{{\rm Pr}\,(C|\Delta  \tilde{\bm
    t})P(\Delta  \tilde{\bm t})}{{\rm Pr}\,(C)}\,.\label{eqs:Bayes theorem}
\end{equation}
In the Poisson process, the arrival times of events are independent and follow
the exponential distribution ${\rm Expo}(r)$, which leads to the joint
distribution
\begin{equation}
    P(\Delta \tilde{\bm t}) = r^n\exp\left(-r\sum_{i=1}^{n} \Delta
    \tilde{t}_i\right)\,.
    \label{eqs:joint distribution of time difference in Poisson process}
\end{equation}
Then, the number of events $N$ in the observation duration $\tau$ follows the
Poisson distribution ${\rm Pois}(r\tau)$, 
\begin{equation}
    {\rm Pr}\,(C) = \frac{r^n\tau^n}{n!}e^{-r\tau}\,.\label{eqs:prob of C}
\end{equation}
For $P(\Delta \tilde{\bm t}|C)$, we find that the constraint $\sum_{i=1}^{n}
\Delta \tilde{t}_i \leq \tau$ naturally arises for ${\rm Pr}\,(C|\Delta
\tilde{\bm t})$ to be non-zero. To ensure that there is exactly $n$ events in
the observation duration $\tau$ given the first $n$ arrival times, the arrival
time of the $(n+1)$-th event should be larger than $\tau-\sum_{i=1}^{n} \Delta
\tilde{t}_i$, so ${\rm Pr}\,(C|\Delta  \tilde{\bm t})$ reads
\begin{equation}
    {\rm Pr}\,(C|\Delta  \tilde{\bm t}) = {\rm Pr}\,\Big(\big\{{\Delta
    \widetilde{T}_{n+1}}>\tau-\sum_{i=1}^{n} \Delta \tilde{t}_i\big\}\Big) =
    \exp\left[-r\left(\tau-\sum_{i=1}^{n} \Delta
    \tilde{t}_i\right)\right]\,.
    \label{eqs:prob of C given time difference}
\end{equation}
Substituting Eqs.~(\ref{eqs:joint distribution of time difference in Poisson
process}--\ref{eqs:prob of C given time difference}) into
Eq.~\eqref{eqs:Bayes theorem}, we obtain
\begin{equation}
    P(\Delta \tilde{\bm t}|C) = \frac{n!}{\tau^n}\,,\ \ \Delta \tilde{t}_i\geq
    0\ \ \& \ \ \sum_{i=1}^{n} \Delta \tilde{t}_i\leq \tau\,,
\end{equation}
which is the same as Eq.~\eqref{eq:joint distribution of dimensionless time
difference} after normalizing with $\tau$. 

\bibliography{rate_refs.bib}
\bibliographystyle{aasjournalv7}

\end{document}